\documentclass{article}
\usepackage{spconf,amsmath,graphicx}
\usepackage{threeparttable} 
\usepackage{booktabs}
\usepackage{multirow}
\usepackage{color}

\usepackage[activate]{microtype}
\newcommand{\eg}{e.\,g.,\ }
\newcommand{\ie}{i.\,e.,\ }

\title{HNAFormer: A Hierarchical Non-Attention Framework for Alzheimer's Detection from Spontaneous Speech}
\title{HAFFormer: A Hierarchical Attention-Free Framework \\for Alzheimer's Disease Detection from Spontaneous Speech}
\name{Zhongren Dong$^1$, Zixing Zhang$^1$*, Weixiang Xu$^1$, Jing Han$^2$*, Jianjun Ou$^3$, Bj\"orn W. Schuller$^4$ \thanks{$^*$ Corresponding authors. zixingzhang@hnu.edu.cn, jh2298@cam.ac.uk}}
\address{$^1$ College of Computer Science and Electronic Engineering, Hunan University, China\\
$^2$ Department of Computer Science and Technology, University of Cambridge, UK \\ 
$^3$ Department of Psychiatry, The Second Xiangya Hospital, Central South University, China \\ 
$^4$ GLAM, Department of Computing, Imperial College London, UK }

\begin{document}
\ninept
\maketitle
\begin{abstract}

Automatically detecting Alzheimer's Disease (AD) from spontaneous speech plays an important role in its early diagnosis. Recent approaches highly rely on the Transformer architectures due to its efficiency in modelling long-range context dependencies. However, the quadratic increase in computational complexity associated with self-attention and the length of audio poses a challenge when deploying such models on edge devices. In this context, we construct a novel framework, namely Hierarchical Attention-Free Transformer (HAFFormer), to better deal with long speech for AD detection. Specifically, we employ an attention-free module of Multi-Scale Depthwise Convolution to replace the self-attention and thus avoid the expensive computation, and a GELU-based Gated Linear Unit to replace the feedforward layer, aiming to automatically filter out the redundant information. Moreover, we design a hierarchical structure to force it to learn a variety of information grains, from the frame level to the dialogue level. By conducting extensive experiments on the ADReSS-M dataset, the introduced HAFFormer can achieve competitive results (82.6\% accuracy) with other recent work, but with significant computational complexity and model size reduction compared to the standard Transformer. This shows the efficiency of HAFFormer in dealing with long audio for AD detection. 


\end{abstract}
\vspace{-.4cm}
\section{Introduction}
\label{sec:intro}
\vspace{-.2cm}
Alzheimer's Disease (AD) is a common neurodegenerative disorder characterised by clinical features, such as memory impairment, language difficulties, executive dysfunction, and cognitive decline. According to the 2020 report of the Lancet Commission~\cite{livingston2020dementia}, there are over 50 million people worldwide affected by dementia in 2020, and this number will be projected to 152 million by 2050. This will substantially increase individual, family, and society's financial burden.
Although AD cannot be completely cured, early screening or diagnosis not only provides more treatment options but also helps slow the progression of the disease and improve the quality of life for patients. However, accurately detecting AD in its early stage is challenging due to the lack of related health knowledge of AD patients. As spontaneous speech is a potential biomarker that relates to the development of AD, automatic speech analysis as method has recently received significant attention, because it offers a non-invasive, cost-effective, and repeatable means to monitor patients' speech characteristics~\cite{gainotti2014neuropsychological, luz2023multilingual}.

Some researchers have focused on AD detection based on spontaneous speech~\cite{balagopalan2021comparing, rohanian2021alzheimer, li2023leveraging}. Rohanian et al.~\cite{rohanian2021alzheimer} integrated rule-based features such as word probabilities, disfluency features, and pause information, with various acoustic features for AD detection. Balagopalan et al.~\cite{balagopalan2021comparing} conducted an in-depth investigation on traditional acoustic features and pre-trained deep features. They found that combining the classic acoustic features with pre-trained acoustic embeddings in a classification approach can yield higher and more robust performance in an unbalanced data distribution. Additionally, Li et al.~\cite{li2023leveraging} explored the impact of acoustic \& linguistic embeddings on AD detection tasks. These methods have primarily concentrated on exploring various speech features, such as paralinguistic features and pre-trained acoustic embeddings. 
Due to the Transformer's capability to model long-range context dependencies, some recent advancements started to employ the Transformer and its variants as classifiers for AD detection~\cite{ilias2023detecting, jin2023consen, mei2023ustc}. Specifically, Ilias et al.~\cite{ilias2023detecting} explored the Vision Transformer (ViT) for AD detection. Jin et al.~\cite{jin2023consen} integrated advanced acoustic embeddings and disfluency features, and combined them with the Swin Transformer and a Random Forest classifier, which achieved the best results in the ADReSS-M competition~\cite{luz2023multilingual}. Mei et al.~\cite{mei2023ustc} also obtained promising results by fine-tuning the wav2vec 2.0 model. 

However, handling long-duration spontaneous speech sequences still remains a challenge. Most of the aforementioned methods rely on standard Transformer architectures, in which the computational complexity of its self-attention module has a quadratic relationship with the input length. This makes the model computationally expensive for long sequences, and thus largely hinders the application of spontaneous speech analysis modelling to the automated AD detection in practice.
Despite that some efforts have been made, such as HTS-AT~\cite{chen2022hts} and HATN~\cite{zhao2020hatn}, to sequentially downsample the audio signals, the original self-attention module retains and thus cannot help relieve the computation issue significantly. 

\begin{figure*}[!h]
  \centering
  \includegraphics{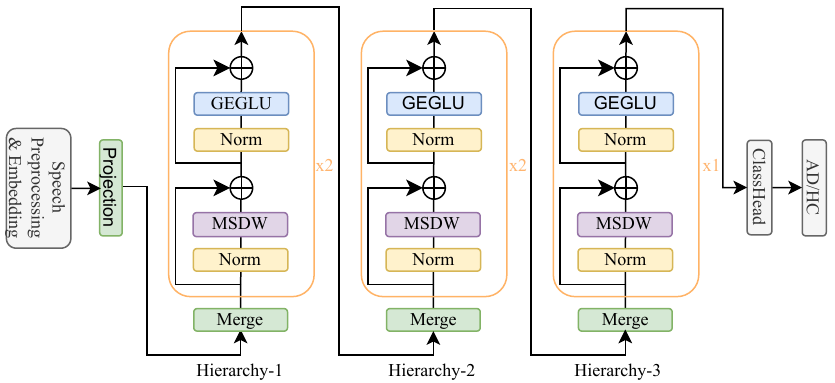}
  \vspace{-.2cm}
  \caption{An overview of the proposed Hierarchical Attention-Free Transformer (HAFFormer) framework for Alzheimer's disease detection. MSDW: multi-scale depth-wise convolution; GEGLU: GELU-based Gated Linear Units; AD/HC: Alzheimer's Disease or Healthy Control.}
  \label{Framework}
  \vspace{-.4cm}
\end{figure*}

To address this challenge, we designed a more efficient architecture, namely the \textit{Hierarchical Attention-Free Transformer} (HAFFormer). This design is partially inspired by the Metaformer~\cite{yu2022metaformer} framework, where the Transformer architecture can be mainly structured with a Token Mixer (equal to the original self-attention module) and a Channel Mixer (equal to the original feedforward module). The proposed HAFFormer abandons the quadratic computational complexity of self-attention and instead introduces a convolutional module as the Token Mixer. This convolutional module utilises effective depthwise convolution operations, resulting in a remarkable reduction of the model size and more importantly the computational complexity. 
Furthermore, we have implemented the GELU-based Gated Linear Unit (GEGLU) 
module as the Channel Mixer to better capture crucial information within the speech signal. Additionally, encouraged by the SpeechFormer++~\cite{chen2023speechformer++}, HAFFormer is constructed in a hierarchical manner, which further reduces the processing cost for long speech data and makes it more efficient for handling spontaneous speech data. Our contributions are as follows:
\begin{itemize}
    \item We propose an  Attention-Free Transformer (AFFormer) 
    modelling without an attention mechanism, which is specifically designed for dealing with long-duration speech audio signals in our AD detection scenario. 
    \item We introduce a hierarchical structure for the AFFormer, empowering the model with the capability to capture different grains of context information, from fine-grain to coarse-grain. 
    \item We conduct extensive experiments on a publicly available dataset of ADReSS-M for AD detection. The experimental results show that HAFFormer is competitive with other SOTA approaches, but with a considerable reduction of model size and computational complexity. 
\end{itemize}

\section{Hierarchical Attention-Free Transformer}
\label{sec:format}
Figure~\ref{Framework} illustrates the overall framework of the proposed hierarchical network framework -- HAFFormer, which consists of four main components: speech preprocessing and embedding, a projection, a hierarchy of merge and AFFormer blocks, and a classification head. Specifically, 
 as shown in Fig.~\ref{Framework}, we introduce three hierarchies of merge and AFFormer blocks. Each of these hierarchies plays a different role in capturing context information, ranging from local to global. 
 The classification head is designed for binary AD detection and is composed of two fully connected layers. In the following sections, we elaborate on the first three components, respectively. 


\vspace{-.4cm}
\subsection{Speech Preprocessing and Embedding}
\vspace{-.1cm}
To obtain rich AD detection features from speech, we employ advanced pretrained models for acoustic feature extraction. Currently, popular pretrained speech models via Self-Supervised Learning (SSL) include Wav2Vec 2.0 \cite{baevski2020wav2vec}, HuBERT \cite{hsu2021hubert}, WavLM \cite{chen2022wavlm}, and others. 
Given that the considered AD dataset is cross-lingual (ref.~Section~\ref{subsec:dataset}) and early versions of Wav2Vec 2.0, HuBERT, and WavLM were trained on English data only, we opt for Wav2Vec2 XLS-R \cite{babu2021xls}. This model shares the same architecture with Wav2Vec 2.0 but was trained on much larger datasets over 128 languages to better capture cross-lingual speech representation.

It is worth noting that the AD data that are spontaneous speech (dialogue) often tend to be pretty long, such as several minutes. However, conventional pre-trained models (\eg  Wav2Vec2 XLS-R) typically accept the maximal speech length of 30 seconds, or even less, as inputs. To address this issue, we segment the entire long speech into sequential short utterances using WhisperX~\cite{bain2023whisperx}, by transcribing the signals into linguistic sentences as well as their temporal boundaries. We then feed the sequential short utterances into the Wav2Vec2 XLS-R model to extract frame-wise speech embeddings, which are then concatenated subsequently. The dimensionality of each speech embedding is denoted as $D = L \times N$, where $N$ is 1024 and $L$ depends on the length of the entire speech signals. In this paper, we set $L$ to be 3200 (64 seconds) which can cover most of the speech samples (ref.~Section~\ref{subsec:dataset}). Any speech longer than 3200 
is truncated, while the ones shorter than 3200
are padded with zeros.


\vspace{-.4cm}
\subsection{Projection}
\vspace{-.2cm}
Considering that AD patients' spontaneous speech is often lengthy and the annotations are scarce, a high representation dimension would often lead to an overfitting problem. Moreover, it often inevitably brings a heavy computational load. To mitigate the risk of overfitting and reduce the computational load for subsequent AFFormer blocks, we introduce a projection layer to map the high-dimensional representations into low-dimensional ones. Assuming a speech signal corresponds to an embedding of $X$ with dimension $D = 3200 \times 1024$, the projection layer is introduced to reduce it to a lower dimensionality of $D = 3200 \times 8$.



\vspace{-.4cm}
\subsection{Merge}
\vspace{-.2cm}
As shown in Fig.~\ref{Framework}, each hierarchy incorporates a merge layer and one/two AFFormer block(s), where the merge layer serves a dual purpose: reducing the computational complexity associated with long data in Transformer-based variant models and Considerably eliminating the inherent redundancy within speech data for feature aggregation.
Specifically, the merge blocks are also implemented by Conv1D. In our experiments, the first, second, and third merge modules downsample the data by factors of 4, 2, and 2, yielding the output dimension of  $D = 800 \times 8$, $400 \times 8$, and $200 \times 8$, respectively. 

\vspace{-.4cm}
\subsection{AFFormer block}
\vspace{-.2cm}
Inspired by Metaformer~\cite{yu2022metaformer}, we seek to design the most suitable Token Mixer and Channel Mixer for our AD detection task, aiming to replace the high-complexity self-attention module in the standard Transformer.

\begin{figure}[t]
\vspace{-.2cm}
  \centering
  \includegraphics{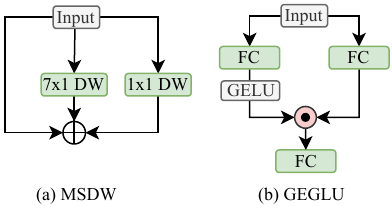}
  \vspace{-.3cm}
  \caption{Detailed architecture of the MSDW module (a) and the GEGLU module (b).}
  \label{GLU}
  \vspace{-.6cm}
\end{figure}

\textbf{Token Mixer:} We design the Token Mixer with the MSDW structure as shown in Fig.~\ref{GLU} (a). 
It consists of a two-branch convolution topology: One branch is a $1 \times 1$ depth-wise convolution; the other one is a $7 \times 1$ depth-wise convolution. Mathematically, the process of Token Mixer can be calculated by:
\vspace{-.2cm}
\begin{equation}
  Y=\text{Conv1D}_{7\times 1}(\text{LN}(X)) + \text{Conv1D}_{1\times 1}(\text{LN}(X)) + X,
  \vspace{-.2cm}
\end{equation}
where $X$ and $Y$ represent the inputs and outputs of the Token Mixer, GELU and LN denote the Gaussian Error Linear Unit activation function and layer normalisation. The module is highly motivated by the  Inverted Separable Convolution (ISC) module, as proposed in MobileNet V2~\cite{sandler2018mobilenetv2}, where the depth-wise convolution (DW) and inverted residual structure are efficient in capturing the local context information while minimising model size.
This design refrains from self-attention layers and thus avoids expensive computations.

\textbf{Channel Mixer:} We construct the Channel Mixer with GEGLU as shown in Fig.~\ref{GLU} (b). The GLU module \cite{dauphin2017language} comprises two branches, where one branch contains one linear layer followed by a nonlinear activation, \ie GELU in our case, as a gating unit, and the other branch is one linear layer only without any activation function. The outputs from the two branches are then combined by using element-wise multiplication. The gating unit can automatically learn to filter the output information, suppressing unimportant information and retaining relevant one. This design is beneficial for the model to learn long-range dependencies. Recently, the GLU module is increasingly being utilised in large language models~\cite{chowdhery2022palm, thoppilan2022lamda} to replace one fully connected (FC) layer in the feedforward (FFN) module as well. The process of Channel Mixer can be expressed by:
\vspace{-.2cm}
\begin{equation}
  Y=(\text{GELU}(\text{LN}(X)W_1)\odot \text{LN}(X)W_{2},
\end{equation}
where $W_1$ and $W_2$ denote the weights of two linear layers, and $X$ and $Y$ represent the inputs and outputs of the Channel Mixer.

\vspace{-.2cm}

\section{Experiments and Results}
\label{sec:pagestyle}
In this section, we will first introduce the used dataset, followed by detailed information about the experimental setup. Finally, we will present the results of the HAFFormer and engage in a discussion.

\vspace{-.4cm}
\subsection{Dataset}
\label{subsec:dataset}
\vspace{-.1cm}
In this study, we utilised the ADReSS-M Challenge dataset \cite{luz2023multilingual}, which is designed for multilingual AD detection through spontaneous speech. The ADReSS-M dataset consists of 291 spontaneous speech samples, where 245 samples are for training and 46 samples for testing. 
In the training set, there are 237 samples in English, and the remaining 8 samples are in Greek. 
The duration of the English training samples ranges from 22.3 seconds to 268.5 seconds, with an average duration of 75.9 seconds. In the testing set, all samples are in Greek, with a duration ranging from 11.8 seconds to 119.4 seconds and an average duration of 38.1 seconds. The training dataset was balanced with respect to age and gender to mitigate potential confounds and biases in the data.


\vspace{-.2cm}
\subsection{Experimental Setup}
\vspace{-.1cm}
In our implementation using the PyTorch framework, we employed the cross-entropy loss function to minimise the loss. We used the AdamW optimiser with an initial learning rate of 2e-3 and a weight decay rate of 1e-5. The batch size was set to 8, and the model was trained for 80 epochs. 
To ensure a fair comparison with other works and assess the overall capability of the model, we use accuracy (ACC) and F1-score (F1) as evaluation metrics. 



\vspace{-.2cm}
\subsection{Results and Discussions}
\vspace{-.1cm}
\subsubsection{Comparison with SOTA Approaches}
\vspace{-.1cm}
To demonstrate the effectiveness of our proposed method, we conducted a comparison with other SOTA approaches. As shown in Table~\ref{sota}, the introduced HAFFormer yields better performance than most previous work~\cite{chen2023cross, mei2023ustc}, where various combinations of acoustic features (ComParE 2016, eGeMAPS, wav2vec2-base) and fine-tuning methods (wav2vec2-large-xlsr-53) for pre-trained models have been attempted \cite{chen2023cross, mei2023ustc}. 

It also achieves competitive performance with the best work in~\cite{Tamm2023cross}, where pre-training combined with fixed-batch transfer learning has been employed. Pre-training is conducted on a subset of English data from the training set, followed by a fine-tuning on the remaining English and Greek data. Their model consists of two fully-connected layers and an attention-pooling layer.
Despite the simplicity, this architecture may lack the scaling capability when increasing the amount of training data.


\vspace{-.3cm}
\begin{table}[!t]
  \caption{Performance comparison between the proposed HAFFormer and other state-of-the-art methods.}
  \label{sota}
  \vspace{0.1cm}
  \begin{threeparttable}
  \resizebox{\columnwidth}{!}{
      \begin{tabular}{ccc}
        \toprule
       Methods     &ACC [\%]  &F1 [\%]   \\
        \midrule
        Pre-train + Mixed-batch transfer learning~\cite{Tamm2023cross}  &82.60  &- \\
        Fine-tuned wav2vec 2~\cite{chen2023cross}  &73.91  &-    \\
        IS10-paralinguistics-compat+SVM~\cite{mei2023ustc}   &69.60  &-    \\
        \midrule
        HAFFormer                                        &82.60  &82.60 \\
        
      \bottomrule
    \end{tabular}}
    \vspace{-.4cm}
\end{threeparttable}
\end{table}

\vspace{-.1cm}
\subsubsection{Selection of the Token Mixers}
\vspace{-.1cm}
\label{Token Mixer}
In our exploration of the MetaFormer architecture for AD detection and the search for a more efficient and effective Token Mixer, we tested six different Token Mixers: Self-Attention, Pool, Identity, ISC, DW, and MSDW, with Channel Mixer fixed with FFN. The relevant parameters were set as follows: d\_model for Self-Attention was set to 8, and for ISC, DW, and MSDW, the number of convolutional kernels was set to 8. The number of neurons in FFN was set as 4 times d\_model, following the standard Transformer setting. Additionally, the $1 \times 1$ convolution in ISC was replaced with an FC layer, using the same parameters as the FFN.

\begin{figure}[t]
  \centering
  \includegraphics[width=.45\textwidth]{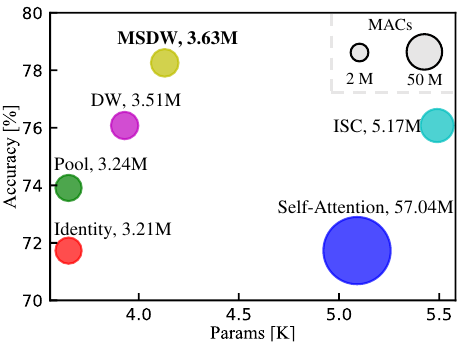}
  \vspace{-.4cm}
  \caption{Model performance (accuracy) vs model size (number of parameters) when taking different types of token mixers under the same channel mixer (\ie FFN). MACs : Multiply–accumulate operations.} 
  \label{token-mixer}
  \vspace{-.6cm}
\end{figure}

The results are shown in Fig.~\ref{token-mixer}. It is illustrated that, with the same Channel Mixer, both Identity and Pool outperformed Self-Attention, demonstrating the effectiveness of the MetaFormer architecture in AD detection. Furthermore, MSDW achieved the best results with
considerably 
fewer parameters than Self-Attention. Therefore, we selected MSDW as the token mixer due to its superior performance while requiring far fewer parameters compared to self-attention.

\vspace{-.2cm}

\vspace{-.1cm}
\subsubsection{Selection of the Channel Mixers}
\label{Channel Mixer}
\vspace{-.1cm}
In this section, we aim to select the best Channel Mixer. We evaluated four different Channel Mixers: FFN, Pool, Identity, and GEGLU. We set the number of neurons in GEGLU as 2 times d\_model, with other parameters set the same as in Section~\ref{Token Mixer}. Table~\ref{channel-mixer} presents the results for different Channel Mixers. The experimental results indicate that GEGLU consistently yields good results, with MSDW + GELU achieving the best performance. Notably, the MACs for MSDW + GELU are approximately $\frac{1}{20}$ of the standard Transformer, making it highly efficient in terms of the number of model parameters.

It is worth mentioning that the parameters listed in Section \ref{Token Mixer} and Section \ref{Channel Mixer} exclude the projection layer's parameters. This is because we map the embeddings to a very low dimension, resulting in the parameters of other modules being much smaller than those of the projection layer. For example, the MACs for Self-Attention + FFN are 107.15M, while the MACs for the projection layer are 78.64M, accounting for 73.4\% of the total MACs. On the other hand, the MACs for MSDW + GELU are 80.08M, with the projection layer's MACs being 78.64M, accounting for over 98\% of the total MACs. Therefore, the parameters listed are without the projection layer, for a fair comparison.

\vspace{-.1cm}
\begin{table}[!t]
  \caption{Comparison of the model performance (accuracy [ACC] and F1), the model size (number of parameters), and the model operation complexity (number of MACs) by taking different token mixers and channel mixers. 
  Pool: average pooling; Identity: no any module; ISC: inverted separable convolution; DW: depth-wise convolution; MSDW: multi-scale depth-wise convolution. FFN: feedforward layer; GEGLU: GELU-based gated linear unit.} \vspace{0.1cm}
  \label{channel-mixer}
  \begin{threeparttable}
  \resizebox{\columnwidth}{!}{
      \begin{tabular}{cccccc}
        \toprule
       Token     &Channel    &ACC [\%]  &F1 [\%]  &Params [K] &MACs [M] \\
        Mixers    & Mixers   &  &  & & \\
        \midrule
        \multirow{4}{*}{Self-Attention} &FFN       &71.73    &71.69    &5.09     &28.51  \\ 
                                        &Pool      &80.43    &80.29    &2.33     &27.18   \\
                                        &Identity  &73.91    &73.91    &2.33     &27.18  \\
                                        &GEGLU     &78.26    &78.01    &4.45     &28.18     \\
        \midrule
        \multirow{4}{*}{Pool}           &FFN       &73.91    &73.91    &3.65     &1.6     \\ 
                                        &Pool      &76.08    &76.08    &0.89     &0.27    \\
                                        &Identity  &73.91    &73.91    &0.89     &0.27    \\
                                        &GEGLU     &76.08    &76.05    &3.01     &1.27    \\
        \midrule                                
        \multirow{4}{*}{Identity}       &FFN       &71.73    &71.69    &3.65     &1.6     \\ 
                                        &Pool      &80.43    &80.29    &0.89     &0.27    \\
                                        &Identity  &73.91    &73.91    &0.89     &0.27    \\
                                        &GEGLU     &78.26    &78.01    &3.01     &1.27    \\ 
        \midrule                                 
        \multirow{4}{*}{ISC}            &FFN       &76.08    &76.08    &5.49     &2.56     \\ 
                                        &Pool      &76.08    &75.68    &2.73     &1.23     \\
                                        &Identity  &80.43    &80.29    &2.73     &1.23     \\
                                        &GEGLU     &80.43    &80.43    &4.85     &2.23     \\ 
        \midrule                               
        \multirow{4}{*}{DW}             &FFN       &76.08    &76.09    &3.93     &1.75     \\ 
                                        &Pool      &71.73    &71.69    &1.17     &0.42     \\
                                        &Identity  &73.91    &73.91    &1.17     &0.42     \\
                                        &GEGLU     &76.08    &76.08    &3.29     &1.42    \\
        \midrule                                
        \multirow{4}{*}{\textbf{MSDW}}  &FFN       &78.26    &78.26    &4.13     &1.77    \\ 
                                        &Pool      &76.08    &76.89    &1.37     &0.44     \\
                                        &Identity  &76.08    &76.05    &1.37     &0.44\\
                                        &\textbf{GEGLU}  &\textbf{82.60}  &\textbf{82.60}  &3.49  &1.44 \\     
      \bottomrule
    \end{tabular}}
    \vspace{-.4cm}
\end{threeparttable}
\end{table}

\vspace{-.2cm}
\subsubsection{Selection of the Number of Hierarchy}
\vspace{-.1cm}
To reduce the computational complexity of the model for long sequence data, a layered model can be an effective approach. Currently, in both the speech \cite{chen2023speechformer++, chen2022hts} and Computer Vision \cite{yu2022metaformer,wang2023repvit} domains, many works employ a four-hierarchy (stage) model. In this section, we conduct an ablation study on the number of layers in the hierarchical model.

\begin{figure}[!t]
  \centering
  \includegraphics[width=.45\textwidth]{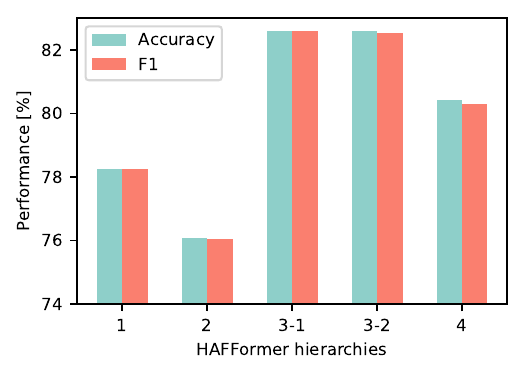}
  \vspace{-.2cm}
  \caption{Performance of the proposed HAFFormer when taking different numbers of hierarchy. 3-1/2 indicates the three hierarchies but the last one contains one or two blocks.}
  \label{stage}
  \vspace{-.3cm}
\end{figure}

From Fig.~\ref{stage}, it can be observed that both Hierarchy3-1 and Hierarchy3-2 achieved the best results. The difference between them lies in the number of layers in the HAFFormer block: Hierarchy3-1 utilises only one layer, while Hierarchy3-2 uses two layers. Consequently, Hierarchy3-1 has fewer parameters. On the other hand, Hierarchy4 experienced a drop in performance due to overfitting, likely caused by an increase in model size relative to the small dataset. This suggests that having a moderate number of layers, as seen in Hierarchy3-1 and Hierarchy3-2, is more effective for the given AD dataset, while increasing the model complexity beyond a certain point may lead to overfitting on this dataset.

\section{CONCLUSION}
\label{sec:typestyle}
To deal with long spontaneous speech in the context of Alzheimer’s disease detection, we proposed a lightweight Transformer variant of Hierarchical Attention-Free Transformer (HAFFormer). The attention-free alternative of the self-attention module 
 remarkably reduces the quadratic computational complexity, and the GELU-based gated linear unit -- an alternative of the feedforward module -- can better learn to select the most salient representation. Moreover, the introduced hierarchical architecture further lowers the processing cost for handling long speech data. The empirical results present the efficiency of the proposed model in terms of model size and complexity for AD detection. The HAFFormer will be further investigated on other related tasks in mobile mental health~\cite{han2021deep}.  



\vfill\pagebreak

\bibliographystyle{IEEEtran}
\bibliography{strings,refs}

\end{document}